\PassOptionsToPackage{table}{xcolor}
\documentclass[sigconf]{acmart}
\AtBeginDocument{%
  }

\setcopyright{acmlicensed}
\copyrightyear{2025}
\acmYear{2025}
\acmDOI{XXXXXXX.XXXXXXX}
\acmConference[EASE '25]{International Conference on Evaluation and Assessment in Software Engineering}{June 17--27,
  2025}{Istanbul, TR}
\acmISBN{978-1-4503-XXXX-X/2018/06}

\usepackage{tikz}
\usepackage{collcell}
\usepackage{xcolor}
\usepackage{etoolbox}
\usepackage{pgf} 
\usepackage{orcidlink}
\definecolor{high}{HTML}{76f013}  
\definecolor{low}{HTML}{ec462e}  

\begin{document}

\title{Incidents During Microservice Decomposition: A Case Study}


\author{Do\u{g}a\c{c} Eldenk 
    \orcidlink{0009-0004-0628-9156}
}
\affiliation{%
  \institution{Carbon Health}
  \city{Evanston}
  \state{Illinois}
  \country{United States}
}
\email{dogac@carbonhealth.com}

\author{H. Alperen \c{C}etin 
\orcidlink{0000-0001-9879-8599}}
\affiliation{%
  \institution{Carbon Health}
  \city{Ankara}
  \country{Turkey}
}
\email{alperen@carbonhealth.com}

\begin{abstract}
    Software errors and incidents are inevitable in web-based applications. Scalability challenges, increasing demand, and ongoing code changes can contribute to such failures. As software architectures evolve rapidly, understanding how and why incidents occur is crucial for enhancing system reliability. In this study, we introduce Carbon Health’s software stack, share our microservices journey, and analyze 107 incidents. Based on these incidents, we share insights and lessons learned on microservice decomposition. Finally, we suggest that starting with monolithic modularization as an initial step toward microservice decomposition may help reduce incidents and contribute to building more resilient software.

\end{abstract}

\ccsdesc[500]{General and reference~Empirical studies}
\ccsdesc[500]{Software and its engineering~Client-server architectures}
\ccsdesc[300]{Software and its engineering~Cloud computing}
\ccsdesc[500]{Software and its engineering~Software post-development issues}
\ccsdesc[500]{Software and its engineering~Software reliability}
\ccsdesc[500]{Software and its engineering~Software fault tolerance}
\keywords{software, incident, outage, microservice, monolith, cloud, distributed, systems, server}

\received{16 March 2025}
\received[revised]{... 2025}
\received[accepted]{... 2025}

\maketitle

\section{Introduction}
Carbon Health is a tech-enabled healthcare company with medical clinics and software to manage electronic health records. Production downtime, bugs, and incidents are notoriously part of every software system \cite{peng1993software, kleppmann2019designing}. Carbon Health's applications are no exception. During the last five years, over 100 incidents were recorded at Carbon Health, while the software architecture of the company has transformed dramatically. 


Software incidents and bugs in cloud and distributed systems have been thoroughly studied. Some studies have examined the causes of system bugs and failures, categorized them, and proposed solutions to mitigate their occurrence \cite{gunawi_what_2014}. Others have focused on detecting software bugs early on to minimize the impact \cite{ganatra_detection_2023, zhang_understanding_2021}. Empirical studies have been conducted on software bugs encountered in various software tools and companies on different scales  \cite{how_bad_openstack, li_empirical_2018}. Additionally, some studies have aimed to make systems more resilient to prevent future incidents \cite{fritzsch_microservices_2019}.



The system transformation process at Carbon Health has primarily involved decomposing a monolith into microservices, which has been a popular topic since the 2010s \cite{kalske2018challenges}. However, it lacks standardized metrics and baselines \cite{abgaz_decomposition_2023}. While the decomposition methods/tools have been extensively studied \cite{martinez_saucedo_migration_2025}, researchers have also explored the motivations and outcomes of the migration using experience reports, case studies, and surveys \cite{lenarduzzi_does_2020, musavi_experience_2016}. On the other hand, some papers have focused on incident mitigation \cite{taibi_processes_2017, velepucha_monoliths_2021}.  Also, challenges faced during the migration have been analyzed using surveys \cite{garrigos_challenges_2018}.

We have found that there is only a small number of studies focusing on cloud-based web applications and analyzing the system-wide incident reports and categorizing their root causes \cite{gunawi_why_2016}. To the best of our knowledge, no study has analyzed system incident causes and patterns concerning the software's architectural evolution, most notably monolith to microservice decomposition.

In this study, we share the details of the software systems at the company, report our experience from different aspects using postmortem incident records, analyze their root causes, and explore how incidents conjunct with the microservice decomposition process.

\section{Background}
This section shares background about software systems, their transformation, and incident management strategies in the company.

\subsection{System Architecture}
Carbon Health's software is accessible via the web and mobile applications used by both care providers and patients. The company has around 100 active clinics and more than a million patients across the US. The software system handles a total of over 2000 requests per second during peak hours. The databases contain over 500 tables with approximately three terabytes of data.

The backend system consists of a monolithic app, 20+ microservices, and a GraphQL service. The monolithic app primarily uses Scala with Play Framework\footnote{\url{https://www.playframework.com/}}, as a RESTful API. Each microservice has a database and is implemented in Kotlin with Armeria framework\footnote{\url{https://armeria.dev/}}, and communicates with other services via gRPC. All the services are hosted in AWS\footnote{\url{htttp://aws.amazon.com/}}. Web and mobile applications share the codebase and are built together using React Native. They directly communicate with the monolith and GraphQL services.

\subsection{System Transformation}



In mid-2021, as the company grew, there were concerns about scalability, slow CI/CD and local build times, slow pull request to release times, and code complexity. Approximately, the number of developers was 10 in 2020, peaked at 100 in 2022, and declined to 30 by 2025. Therefore, with the increasing developer count, a company decision has been made to move our features, primarily starting from the newly built ones, to microservices.


Some of the features have been moved away from the RESTful monolithic app to microservices using gRPC and a GraphQL gateway. 20+ functional microservices have been created in a span of 2 years. Currently, around three-fourths of tables belong to the monolith, and the rest are distributed to microservices. To date, the majority of the features are still in the monolith. 



Our process for moving a feature to a microservice looks like:

\begin{enumerate}
    \item Identify the domain, features, and database tables to migrate.
    \item Write a proposal document to describe the database schema, introduce the new endpoints, and analyze required database queries.
    \item Get reviews from team members and engineering managers. Address reviewers' concerns and agree on the proposal.
    \item Refactor monolith to collect all database operations in the repository layer. Modularize the code if needed. 
    \item Implement the functionality in the microservice.
    \item Start migrating the data either by calling microservice endpoints in batch jobs or using a database replication tool.
    \item Roll out the new microservice to serve production data step by step. Read from both old and new implementations, compare results in the code, and handle bugs.
    \item Clean up the old implementation, database tables, and code from the monolith service after the roll-out reaches 100\% and the app is stable.
\end{enumerate}


As mentioned in step 4 above, we intentionally encapsulated all database access within \textit{repository} classes. By utilizing this pattern, we made the microservice decomposition easier by restricting access to tables and rows during the data migration. Without it, we had to inspect every database query to ensure our tables are not accessed in other queries, such as joins and methods in different repositories, which could lead to data inconsistencies during the data migration process.

After migrating a few microservices, we also started to modularize the monolithic code base. We moved different domains to different packages. Modularization also helped with build times and the separation of concerns. Modules still exposed the repositories and services inside them over interfaces; however, their implementations stayed package private. This has been possible using dependency injection techniques, the resulting code structure was similar to SOA and dynamic binding, which has been used in production systems in the early 2000s \cite{michlmayr_towards_nodate}.

\vspace{-2mm}
\subsection{Incident Management}
At Carbon Health, logs and application metrics are collected from servers and clients. All services are required to expose OpenTelemetry\footnote{\url{https://opentelemetry.io/docs/what-is-opentelemetry/}} metrics, and they are deployed behind an envoy sidecar proxy, which helps us collect low-level network metrics\footnote{System-wide metrics are collected via a central Prometheus collector}. Alerts regarding overall-system health are set using metrics such as endpoint error rate, latency, CPU usage, and JVM garbage collection. Developers add their custom alerts based on business logic. Also, alerts for infrastructure are in place using the tools provided by the cloud providers and CI/CD using notification channels.  

Beyond automated alerts, an on-call software engineer always monitors the system and responds to alerts and notifications. The on-call developers change in a recurring schedule. Generally, the on-call engineer follows a few channels, such as the channel that streams support tickets escalated up to engineering. An issue can be anything, such as a bug affecting a small piece of the software or a full system down. For an issue to be categorized as an incident, it should affect a large number of users, cause a widespread system slowdown/failure, or make a significant part of the system dysfunctional.

Our standard documented process of incident management is as follows. The first step is to mitigate effects such as rolling back to an older version or killing the bad database query. The second step is to find the responsible team and reach out to them to take over the issue and find a solution. After the issue is resolved, the responsible team prepares a postmortem document containing the duration and timeline of the incident, its effect scope, and the whole investigation report on why, what, and how. After the document review process, the on-call engineer and the owner team schedule a meeting with all engineers to explain the issue, educate others, and ask their opinions on the action items. In February 2024, the company began sharing the system health details in the status page\footnote{\url{https://status.carbonhealth.com/}}.


\vspace{-2mm}

\section{Data Collection}

We inspected  107 incident reports from June 2020 to February 2025, then identified their severity in three levels: low, medium, and high. Low severity incidents can be considered as serious bugs, as they only affect a very small set of users and non-critical features. Medium severity incidents affect the majority of users, the system is partially operational; however, critical features might be degraded. High severity incidents affect all users and cause system-wide failure or failure on critical paths. The number of high, medium, and low severity incidents is 65, 29, and 13, respectively. 

Also, the number of incidents resolved in 0-1, 1-3, 3-24, and over 24 hours is 57, 26, 14, and 8, respectively. There are 2 incidents with unknown time to resolution. When developers realized the incidents, it was not possible to detect the beginning, so they reported it as unknown.


Later, we categorized the incidents using an open coding approach. First, we extracted several keywords or phrases that best described each incident. Then, we reviewed all the expressions and grouped them into ten categories:

\begin{enumerate}
    \item Frontend: Failures originating from the frontend code, such as wrong API calls, incorrect data shape, race conditions, and display bugs.
    \item Breaking change: Unintentional breaking changes that broke the compatibility between services and web/mobile clients. 
    \item Refactoring: Unintended behavior during microservice migration, code refactoring, moving stuff around. 
    \item Database: Row/table locking issues, deadlocks, slow queries, incompatible row format with application code.
    \item Over-fetching: Trying to fetch too much data, usually more than required.
    \item Infrastructure: Infrastructure reliability, services being down, scalability issues, issues that happened during provisioning or moving software infrastructure.
    \item CI/CD: Unintended behavior in deployment and build pipelines, such as race conditions, the pipeline being entirely frozen.
    \item Framework: Issues caused by libraries, garbage collector issues, missing updates, and bugs faced while trying to update libraries and frameworks.
    \item Null pointer exception (NPE): Bugs caused by missing data, often due to eventual consistency.
    \item External: Other services being down, such as external APIs.
\end{enumerate}

Table \ref{tab:incident_categories} shows the number of incidents for each category per year. Note that one incident can fall under multiple categories.

\vspace{-2mm}

\section{Discussion}

When looking at the data in Table \ref{tab:incident_categories}, the trends in categories have been different, even though nearly all categories peaked in 2022, the year we worked on decomposition the most.

\textbf{Database:} Following the start of the microservice decomposition in 2021, there has been a decline in \textit{database} incidents and an increase in \textit{over-fetching} and \textit{refactoring} incidents. A deep dive into the incident reports revealed that most issues in the database were primarily caused by excessive locks, reading/filtering large amounts of data, and poor data/query design. These problematic database queries were written when the company had only a few clinics. They quickly hit the limits when the scale increased.

\textbf{Modularization:} After we started refactoring join queries into repository calls in the monolith and banning joins between different domains, performance issues shifted from the database to application code. We started to see unexpected computations depleting our server resources and causing scalability issues. Therefore, the process of modularization surfaced some issues, such as missing indices or poorly designed queries. The decrease in \textit{database} incidents and increase in \textit{refactoring} incidents after 2021 support these findings. The \textit{over-fetching} incidents have increased in 2021, peaked in 2022, and then started disappearing. We believe that modularization encouraged developers to write more efficient and well-tested code, and our data hints that modularization might be more effective than the microservice decomposition in resolving the \textit{over-fetching} issues.


\textbf{Data Migration:} We also gathered experience migrating microservices to both relational databases and non-relational databases during the process. The migration process for relational databases was simpler because the table schema was preserved or modified slightly. However, for non-relational databases, the process for migrating the data was more challenging. For relational databases, we could use joins for tables that belong to the same domain to address performance concerns; however, we could not do that in non-relational databases. Also, services migrated to non-relational databases had more data, hundreds of millions of rows, compared to a few million rows in relational databases. The back-filling and testing process almost took two months for the biggest table in the company, and caused multiple performance and \textit{over-fetching} incidents. Moreover, switching to non-relational databases produced unique types of incidents caused by the eventual consistency of data, as most of the application code was written around strong consistency assumptions. We merged these incidents under the \textit{NPE} category.

\begin{table}[b]
    \centering
    \vspace{-6mm}
    \caption{Number of incidents by category and year}
    \vspace{-3mm}
    \begin{tabular}{lccccccc}
        \toprule
        & 2020 & 2021 & 2022 & 2023 & 2024 & 2025 & Total  \\
        \midrule
        Database & \cellcolor{red!4} 2 & \cellcolor{red!28} 14 & \cellcolor{red!18} 9 & \cellcolor{red!8} 4 & \cellcolor{red!0} 0 & \cellcolor{red!2} 1 & \cellcolor{red!60} 30 \\
        Refactoring & \cellcolor{red!0} 0 & \cellcolor{red!0} 0 & \cellcolor{red!26} 13 & \cellcolor{red!14} 7 & \cellcolor{red!6} 3 & \cellcolor{red!0} 0 & \cellcolor{red!46} 23 \\
        Over-fetching & \cellcolor{red!2} 1 & \cellcolor{red!14} 7 & \cellcolor{red!22} 11 & \cellcolor{red!8} 4 & \cellcolor{red!0} 0 & \cellcolor{red!0} 0 & \cellcolor{red!46} 23 \\
        Infrastructure & \cellcolor{red!0} 0 & \cellcolor{red!8} 4 & \cellcolor{red!22} 11 & \cellcolor{red!6} 3 & \cellcolor{red!6} 3 & \cellcolor{red!0} 0 & \cellcolor{red!42} 21 \\
        Breaking cha. & \cellcolor{red!2} 1 & \cellcolor{red!8} 4 & \cellcolor{red!8} 4 & \cellcolor{red!6} 3 & \cellcolor{red!8} 4 & \cellcolor{red!2} 1 & \cellcolor{red!34} 17 \\
        Frontend & \cellcolor{red!4} 2 & \cellcolor{red!8} 4 & \cellcolor{red!12} 6 & \cellcolor{red!2} 1 & \cellcolor{red!8} 4 & \cellcolor{red!0} 0 & \cellcolor{red!34} 17 \\
        CI/CD & \cellcolor{red!0} 0 & \cellcolor{red!2} 1 & \cellcolor{red!14} 7 & \cellcolor{red!6} 3 & \cellcolor{red!0} 0 & \cellcolor{red!0} 0 & \cellcolor{red!22} 11 \\
        External & \cellcolor{red!0} 0 & \cellcolor{red!6} 3 & \cellcolor{red!10} 5 & \cellcolor{red!0} 0 & \cellcolor{red!6} 3 & \cellcolor{red!0} 0 & \cellcolor{red!22} 11 \\
        Framework & \cellcolor{red!0} 0 & \cellcolor{red!2} 1 & \cellcolor{red!12} 6 & \cellcolor{red!2} 1 & \cellcolor{red!0} 0 & \cellcolor{red!0} 0 & \cellcolor{red!16} 8 \\
        NPE & \cellcolor{red!2} 1 & \cellcolor{red!2} 1 & \cellcolor{red!4} 2 & \cellcolor{red!4} 2 & \cellcolor{red!0} 0 & \cellcolor{red!0} 0 & \cellcolor{red!12} 6 \\
        \bottomrule
    \end{tabular}
    \label{tab:incident_categories}
\end{table}

\textbf{New Technologies:} We started encountering a new set of issues related to microservices after completing repository decomposition, such as network-level request limits, network partition failures, and unexpected scaling issues with new frameworks related to threading. These issues took time and expertise to understand and resolve. This aligns with the increase in \textit{framework} incidents in 2022. We observed that documentation was a very important factor in preventing those issues from happening again. However, multiple teams encountered these issues at different times, as teams typically operated autonomously. We believe cross-team collaborations could have prevented the same incident from happening in different domains.

\textbf{Platforming}: Later, company-wide measures have been implemented to prevent poorly performing code from impacting overall system stability. The platform team started publishing a catalog of libraries and tools that are recommended for use by every team to build services. A common issue was writing blocking code and executing it in a non-blocking thread, such as a Netty event loop, where the number of threads is low and close to the number of CPU cores. The blocking executor has thousands of redundant threads, unlike the event loop, which has only two threads per CPU core. Executing all code in a blocking event executor by default improved stability significantly while sacrificing performance slightly. No major issues related to thread starvation have happened since then. To prevent such \textit{framework} incidents, we considered our developer behavior while building our platform.


\textbf{Infrastructure:} Similar to frameworks, we also started migrating to new cloud infrastructure and CI/CD technologies during the first phase of the microservice decomposition, as the microservice architecture introduced new requirements.
However, the deployment pipeline remained largely unchanged, consistently following a rolling update approach. During the migration, the only addition was enforcing deployment to the alpha environment before production. To prevent code duplication and accelerate the development of new services, we created a git repository for reusable GitHub actions and Terraform modules to define infrastructure. Building stable libraries and educating ourselves about these technologies took time. Issues happened during the process; however, they were eventually resolved. The incidents in the \textit{infrastructure} and \textit{CI/CD} categories began decreasing after 2022, as shown in Table \ref{tab:incident_categories}.

\textit{Breaking change} and \textit{frontend} incidents have always been present. We could not establish a correlation with microservice decomposition, as their trends stayed relatively the same.



After a long decomposition process, having the required tooling and expertise helped decrease the number of incidents and have a reliable and stable system. From our experience, the most significant factor was the refactoring of the monolith into a modular structure. It helped developers understand the domains, optimize code and queries where required, and many more, as mentioned above. Our experience aligns with the recent popular modular monoliths trend \cite{su_modular_2024}, and we agree that they are a solid middle ground between microservices and monoliths \cite{tsechelidis_modular_2023, goncalves_monolith_2021}.

\vspace{-2mm}

\section{Threats to Validity}
The quality of data drops after 2024 due to a shift in incident reporting methodology, as they mostly do not include a detailed analysis on how, why, and what. Also, during 2021, the COVID pandemic peaked, and our system was under heavy load as we were primarily working on vaccination and testing. The number of developers has changed over time. Incident data might correlate with these factors. Additionally, our results could be affected by the limited data due to unavailable historical system metrics and privacy constraints. 


\vspace{-2mm}

\section{Conclusion and Future Work}

During Carbon Health’s rapid growth and microservice decomposition journey, numerous incidents were experienced. First, we presented the system architecture, decomposition process, and incident management workflow. Then, we shared the challenges we faced, along with our insights and experiences, based on 107 software incidents over five years. We believe modularization and expertise in the technologies used played a crucial role in preventing incidents and minimizing their impact.


Confirming our results by asking a broader community of developers via questionnaires could be a future direction. Another future research area could be sharing more experiences on modular monoliths, since they might become more popular in upcoming years, as hardware improves and becomes more capable of running faster servers, compilers, language servers, and development tools.



\vspace{-2mm}

\section*{Data Availability Statement}

The incident data and our postmortem document template are publicly available at \url{https://github.com/Dogacel/ease-2025}.

\vspace{-1mm}

\bibliographystyle{ACM-Reference-Format}
\bibliography{references}

\end{document}